\documentclass[aps,pre,twocolumn,amsmath,10pt,groupedaddress,superscriptaddress,amssymb]{revtex4-1}

\usepackage{graphicx}   
\usepackage{subfigure}  
\usepackage{hyperref}  

\newcommand{\ouN}{O\left(\frac{1}{N}\right)}
\newcommand{\dd}{\mathrm{d}}

\newcommand{\s}{\sigma}
\newcommand{\be}{\begin{equation}}
\newcommand{\ee}{\end{equation}}

\DeclareMathOperator*{\tr}{Tr}

\begin{document}

\title{Finite size corrections to disordered Ising models on Random Regular Graphs}

\author{C. Lucibello}
\affiliation{Dipartimento di Fisica,``Sapienza'' University of Rome, P.le A. Moro 2, I-00185, Rome, Italy}
\author{F. Morone}
\affiliation{Levich Institute and Physics Department, City College of New York, New York, New York 10031, USA}
\author{G. Parisi}
\affiliation{Dipartimento di Fisica, IPCF-CNR, UOS Roma, and INFN, Sezione di Roma1, ``Sapienza'' University of Rome, P.le A. Moro 2, I-00185, Rome, Italy}
\author{F. Ricci-Tersenghi}
\affiliation{Dipartimento di Fisica, IPCF-CNR, UOS Roma, and INFN, Sezione di Roma1, ``Sapienza'' University of Rome, P.le A. Moro 2, I-00185, Rome, Italy}
\author{Tommaso Rizzo}
\affiliation{IPCF-CNR, UOS Roma, ``Sapienza'' University of Rome, P.le A. Moro 2, I-00185, Rome, Italy}

\begin{abstract}
We derive the analytical expression for the first finite size correction to the average  free energy of disordered Ising models on random regular graphs. The formula can be physically interpreted as a weighted sum  over all non self-intersecting loops  in the graph, the weight being the free-energy shift due to the addition of the loop to an infinite tree.
\end{abstract}

\maketitle

\section{Introduction}
The investigation of statistical mechanics systems with quenched  disorder, such as spin glasses and random field models, has challenged theoretical physicists as well as mathematicians for many years.  The heuristic and  rigorous techniques developed in this field, namely the replica and cavity methods \cite{Parisi1987,Montanari2009}, the interpolation technique \cite{Guerra2003,Talagrand2003} and the objective method  \cite{Aldous2003}, proved to have a wide range of applicability and enlightened a very rich phenomenology.

Among mean fields models, much has been established regarding fully-connected topologies, while diluted systems have been an harder problem to tackle. They have been the subject of intensive investigation in the last two decades \cite{Montanari2009} and show some of the properties of finite dimensional models (e.g. the existence of many solutions to the saddle point equation in the Random Field Ising Model (RFIM) \cite{Krzakala2010}, causing the breakdown of dimensional reduction in finite dimension).
The main feature of diluted random graphs, that of being locally tree-like in the large graph limit, is exploited by the cavity method, also known in its simplest (replica symmetric) form as Bethe approximation, to produce exact asymptotic results \cite{Me2001,Montanari2009}.

When dealing with finite dimensional systems though, the presence of many short loops poses huge problems to such analytic techniques and few solid results have been achieved.
Fundamental topics such has the presence of a glassy phase in low-dimensional spin glasses \cite{Janus12,Larson2013} or the value of the critical dimension marking the  breakdown of dimensional reduction in the RFIM \cite{Tissier2011} are still at the centre of a much heated debate. 

The development of a perturbative formalism around the Bethe approximation, to systematically include the effect of loops in a graph, is highly desirable and could shed  some light on these problems. This task has been undertaken in the last years using different approaches \cite{Montanari2005,Chertkov2006,Chertkov2006a,Parisi2006,Vontobel2013,Mori2012} but still the computations remain analytically and computationally challenging.

Here we  focus on disordered systems with a random topology, the one of random regular graphs (RRG), where the density of finite loops goes to zero as the system size goes to infinity. In the thermodynamic limit the free energy density of the system can be described through the cavity method (or equivalently the replica method), being the one of a Bethe lattice. 
When the number of vertices $N$ in the graph is finite though, the average free energy density $f(N)$  resents the presence of loops. If $f(N)$ has a regular expansion around $N=\infty$, each term of the $1/N$ expansion $f(N)=f_0+f_1/N +o(1/N)$ would account for the contribution of a certain class of loopy structures. We see that in the context of diluted systems, finite size corrections and loop expansions are strictly related concepts.
In this paper we set up a formalism, based on a replicated action, apt to the systematic computation of the $f(N)$ expansion for disordered Ising systems in the replica symmetric phase. We calculate explicitly the first correction $f_1$ to the thermodynamic free energy. 
It is simple combinatorics to show that only simple (i.e. non-intersecting) loops can participate to the $O(1/N)$ correction $f_1$. In fact more complicated loopy subgraphs, with none dangling nodes, typically involve only a  fraction $O(1/N^2)$ of the total number of nodes, therefore can only contribute to higher order terms in the free energy expansion. Obviously this fact naturally emerges from the analytic computation as well.

In Section \ref{sec:repl} we introduce the replicated formalism for the RRGs and compute the saddle point approximation, the details of the calculation being left to Appendix \ref{app:App1}. In Section \ref{sec:finite} and Appendices \ref{app:App2} and \ref{app:tr}, we compute the $O(1/N)$ finite size correction to the average free energy  from the Gaussian fluctuations around the saddle point. It is given by the formula
\be
f_1 = \sum_{\ell=3}^{\infty} \mathcal{N}(\ell) \,\Delta\phi_\ell ,
\label{eqf1}
\ee
where $\mathcal{N}(\ell)$ is the average number of loops of length $\ell$ in the graph, and $\Delta\phi_\ell$ is the free-energy shift given by the addition of a non-intersecting loops of length $\ell$ to an infinite tree. A similar though slightly different result was found by the authors in the context of Erd\"{o}s-R\'enyi (ER) random graphs \cite{Ferrari2013}. It the ER case
the formula
\begin{equation}
f^{\text{ER}}_1=\ \phi^{\text{ER}} +\sum_{\ell=3}^{\infty} \mathcal{N}(\ell) \,\Delta\phi^{\text{ER}}_\ell ,
\label{eqf1ER}
\end{equation}
there is an additional term, $\phi^{\text{ER}}$, containing the free energies of open chains of length $\ell=0,1,2$.
This additional term is related to the fluctuations in the nodes' connectivity
 and it is absent on the RRGs.

 In Section \ref{sec:cavity} we rederive Eq. \eqref{eqf1} and clarify its meaning using a simple probabilistic (cavity) argument.

We note that  the coefficients $\Delta\phi_\ell$ remain exactly the same in the $O(1/M)$ term of the free energy  expansion suggested in Ref. \cite{Vontobel2013,Mori2012} for finite dimensional systems, only the combinatorial factor $\mathcal{N}(\ell)$ changes accordingly to the number of non-backtracking paths present in the lattice.

To test the analytical result, we performed a numerical experiment on a $\pm J$ spin glass in a uniform external field $H$, and we found a good agreement between the theory and the simulations for different values of $H$, as reported in Section \ref{sec:num}.

\section{Replica formalism for random regular graphs}
\label{sec:repl}
We consider a model constituted by $N$ interacting Ising spins $\s_i=\pm 1$, $i=1,\dots,N$, defined by the Hamiltonian
\be
\mathcal{H}=-\sum_{i<j}C_{ij}\s_iJ_{ij}\s_j-\sum_{i}H_i\s_i ,
\ee
where the exchange couplings $J_{ij}$ and/or the local magnetic fields $H_i$ are  quenched independent random variables. The numbers $C_{ij}$ represent the entries of the adjacency matrix of a graph $G$ extracted from the Random Regular Graph (RRG) ensemble. They take the values $C_{ij}=1,0$ depending on whether or not  the vertices $i$ and $j$ are connected. Here we study $c$-RRGs, i.e., random graphs with vertices  having uniform degree $c$. The probability measure of the $c$-RRG ensemble is uniform over all the regular graphs of degree $c$. In order to compute the average free energy density of this model we use the replica trick \cite{Parisi1987}, that is we exploit the limit
\be
-\beta N f(\beta,N)=\lim_{n\to 0}\ \partial_n\log[Z^n(\beta)]_{\mathrm{av}} ,
\label{eq:replicatedF}
\ee
where $\beta$ is the inverse temperature of the system and $[\,\bullet\,]_{\mathrm{av}}$ denotes the average over the topological disorder (i.e. over the $c$-RRG ensemble) and the random couplings  $J_{ij}$ and fields $H_i$. In the following we will omit the explicit dependence of $Z(\beta)$ and other quantities from $\beta$.

As usual in the replica trick, we compute the integer moments $[Z^n]_{\mathrm{av}}$ of the partition function and continue analytically the resulting expression to real and small values of the replica number $n$, as needed by Eq. \eqref{eq:replicatedF}.

To solve the model in the thermodynamic limit (i.e. $N\to\infty$) and to obtain the finite size correction to this limit, we have to cast the averaged replicated partition function $[Z^n]_{\mathrm{av}}$ into an integral form suited to steepest descent evaluation. This procedure uses standard techniques \cite{Viana1985, *DeDominicis1987, *Mezard1987b, *Kanter1987} and is report in detail in the Appendix \ref{app:App1}. Here we quote only the final result, which reads
\be
[Z^n]_{\mathrm{av}} = [\det(cU)]^{1/2} e^{\mathcal{A}(N,c)} 
\int\mathcal{D}\rho\ e^{-N\mathcal{S}[\rho,N]} ,
\label{eq:IntegralReprOfZ}
\ee
where the meaning of the different terms is presented below, exclusion made for 
the explicit expression of the constant $\mathcal{A}(N,c)$, whose definition can be found in the Appendix \ref{app:App1}.
In the last equation the integral is performed over the space of all possible complex-valued functions $\rho(\s)\equiv\rho(\s_1,\dots,\s_n)$ of a $n$-replicated spin, taking $2^n$ different values. The action $\mathcal{S}[\rho,N]$ is a functional of $\rho(\s)$ and $N$, and at the leading order in $N$ can be written as
\begin{equation}
\begin{aligned}
\mathcal{S}_0[\rho]\,=&\frac{c}{2}\int\dd\s\dd\tau\ \rho(\s)U(\s,\tau)\rho(\tau)\\
&-\log\int\dd\s\ e^{B(\s)}\left[
\int\dd\tau U(\s,\tau)\rho(\tau)
\right]^c
\end{aligned}
\label{S0}
\end{equation}

The action $S_0$ will be optimized through the steepest-descent method. After that, we will integrate the Gaussian fluctuations around the optimal saddle point, thus obtaining the desired finite size corrections. The quantities $U(\s,\tau)$ and $B(\s)$  appearing in Eq. \eqref{S0} are defined  by 
\be
\begin{aligned}
U(\s,\tau)=&\,
 \mathbb{E}_J
\left[
\exp\left(\beta J\sum_{a=1}^n\s^a\tau^a
\right)
\right]
\end{aligned}
\ee
and
\be
\begin{aligned}
B(\s)=\log\mathbb{E}_H
\left[
\exp\left(
\beta H\sum_{a=1}^n\s^a
\right)
\right]\
\end{aligned}
\ee

Saddle point evaluation of $\mathcal{S}_0$ leads to the following self-consistence equation for the order parameter $\rho$:
\be
\rho_*(\s)=\frac{\mathrm{e}^{B(\s)}
\left[
\int\dd\s' U(\s,\s')\rho_*(\s')
\right]^{c-1}
}
{\int\dd\s'' \mathrm{e}^{B(\s'')}
\left[
\int\dd\s' U(\s'',\s')\rho_*(\s')
\right]^{c} 
}.
\label{eq:SPequation}
\ee
Once a solution of Eq. \eqref{eq:SPequation} has been found,  using Eq. \eqref{S0} one gets the thermodynamic free energy density $f_0=\lim_{N\to\infty}f(N)$. The difficulties of the problem are all hidden in the solution $\rho_*(\s)$ of the saddle point equation. The function $\rho_*(\s)$ depends on the replicated spin $(\s_1,\dots,\s_n)$, and hence, it is uniquely determined by the set of the $2^n$ possible values it can take. The most general solution should specify all these $2^n$ values. Here we limit ourselves to the simplest solution, i.e., the replica symmetric one. This solution has the property to be invariant under the group of permutations of the replica indexes, therefore $\rho_*(\s)$ can depend only on the sum  $\sum_{a=1}^{n}\s_a $. The number of parameters necessary to fully specify a replica symmetric function is $n+1$ (and hence much smaller than $2^n$). The most general replica symmetric parametrization of $\rho_*(\s)$ can be written in the  form:
\be
\rho(\s)=\int\dd h\  P(h) \frac{\mathrm{e}^{\beta h\sum_{a=1}^n\s_a}}{[2\cosh(\beta h)]^n} ,
\label{eq:RsParametrization}
\ee
where the function $P(h)$ depends implicitly on $n$ and  is non-negative and normalized to one in the limit $n\to0$. 

Inserting the parametrization \eqref{eq:RsParametrization} into the saddle point equation \eqref{eq:SPequation}, and performing the limit $n\to0$, we obtain a self consistent equation for the density $P(h)$:
\be
P(h)=\mathbb{E}_{J,H}\int\prod_{k=1}^{c-1}\dd h_k\, P(h_k)\, \delta
\left[h-H-\sum_{k=1}^{c-1}\hat{u}(\beta,J,h_k)\right],
\label{eq:cavityEq}
\ee
where $\hat{u}(\beta,x,y)=\beta^{-1}\tanh^{-1}[\tanh(\beta x)\tanh(\beta y)]$. We recognize Eq. \eqref{eq:cavityEq} as the self-consistent equation for the probability distribution $P(h)$ of the
cavity field on a RRG of connectivity $c$. Solving the last equation for $P(h)$ one can eventually evaluate the $n=0$ limit of Eq. \eqref{S0} and recover the thermodynamic free energy $f_0 \equiv\lim_{N\to\infty}\frac{f(N)}{N}$, given by the Bethe free energy approximation \cite{Montanari2009}.

\section{Finite size corrections}
\label{sec:finite}
In this section we present the analytical expression of the first finite size corrections to the free energy density of disordered Ising models on the RRG ensemble. 
As we anticipated in the introduction, we assume the leading correction to the thermodynamical free energy density to be proportional to $1/N$. Therefore, we split $f(N)$ into the sum of the leading term plus the $1/N$ correction, that is 
\be
f(N) = f_0 + \frac{ f_1}{N} + o\left(\frac{1}{N}\right) .
\ee

The detailed calculation of the coefficient $ f_1$, the main result of this paper, is given in the Appendices \ref{app:App2} and \ref{app:tr}. The derivation is based on the expansion of the contributions of the  Gaussian fluctuations of the replicated action around the saddle point, given by
\begin{equation}
-\frac{1}{2}\log \det \left(\frac{\partial^2 S_0}{\partial \rho(\s)\partial \rho(\tau)}\bigg|_{\rho_*}\right),
\end{equation}
as a power series containing the replicated transfer matrix of the system
\cite{Monasson96,Lucibello2014}. The final result reads
\be
f_1=\sum_{\ell=3}^{\infty}\frac{(c-1)^\ell}{2\ell} \Delta\phi_\ell.
\label{eq:Fsc_finalFormula}
\ee
The terms appearing in last equation and computed in the replica formalism have a clear physical meaning, as we will readily explain. We call $\Delta\phi_\ell$ the quantity defined by
\be
\Delta\phi_\ell=\phi_{\ell}^c-\ell\,\phi.
\label{eq:FreeEnergyShift}
\ee
where  $\phi_{\ell}^c$ is the average free energy of a closed chain (loop) of length $\ell$ embedded in the graph, that is 
\begin{equation}
\phi_\ell^c \equiv-\frac{1}{\beta}\big[\log Z^c_\ell\big]_{\mathrm{av}}
\end{equation} 
with 
\begin{equation}
\begin{aligned}
Z^c_\ell&\equiv\sum_{\s_1,\dots,\s_\ell} e^{\beta(r_1 \s_1+J_1\s_1\s_2+\dots +r_\ell \s_\ell+J_\ell\s_\ell\s_1)}.
\end{aligned}
\end{equation}
The cavity fields $r_i$ are i.i.d. random variables sampled from the distribution
\be
R(r)=\mathbb{E}_{J,H}\int\prod_{k=1}^{c-2}\dd h_k\, P(h_k)\,\delta
\left[r-H-\sum_{k=1}^{c-2}\hat{u}(\beta,J,h_k)\right].
\label{eq:cavityEq_for_r}
\ee
In other words, the cavity fields $r_i$ represent the effective fields coming from the rest of the graph on the nodes in a loop. The quantity $\phi$ is the intensive average free energy  of a closed chain with random couplings $J_i$, and random fields $r_i$, i.e. $\phi\equiv \lim_{\ell\to\infty}\frac{\phi_\ell^c}{\ell}$, and can be easily computed through cavity method \cite{Monasson96,Lucibello2014}. 

The fact that the fields $r_i$ are independently distributed and that they obey Eq. \ref{eq:cavityEq_for_r}, containing the fixed point distribution $P(h)$, indicates that the contribution of each loop  can be considered independently from the others. In fact the factor $(c-1)^\ell/2\ell$ in Eq. \eqref{eq:Fsc_finalFormula} is exactly the average number of loops of length $\ell$ in a RRG of connectivity $c$. Therefore, the coefficient $ f_1$ of the $\ouN$ correction can be expressed as a sum over all the loops in a graph, each one contributing with the amount $\Delta\phi_\ell$ to the free energy. We call $\Delta\phi_\ell$ a free energy shift since it is the free energy difference observes in a infinite tree after the addition of a single loop of size $\ell$, as we will argue in the next Section.

It is yet to be investigated the relation between  \eqref{eq:Fsc_finalFormula} for $f_1$ and an analogous result that one could  derive using the loop calculus formalism \cite{Chertkov2006, Chertkov2006a}.

We notice that the loops considered here are  defined as non-self intersecting closed paths. In fact, self-intersecting loops would give a contribution of order  $O(1/N^2)$ to the average free energy for simple combinatorial arguments.

\section{Probabilistic argument}
\label{sec:cavity}
\begin{figure*}
\includegraphics[width=\textwidth]{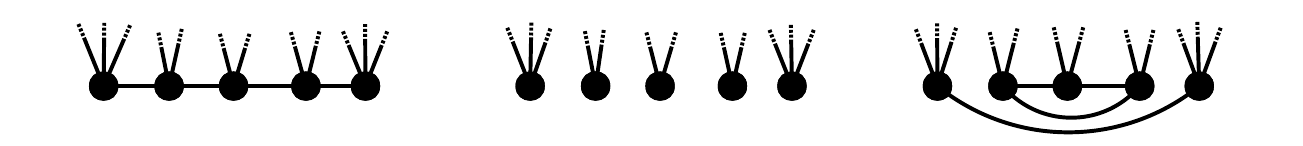}
\caption{Pictorial representation of the argument, given in Section \ref{sec:cavity}, to compute the free energy shift due to the addition of a loop to a large tree graph. It is shown an open chain embedded in a tree graph (\textit{left}), its removal from the tree (\textit{center}) and the addition of a loop (\textit{right}).}
\label{fig:cavity}
\end{figure*}
The computation of the $O(1/N)$ correction to the free energy in the RRG ensemble can be easily done  through simple probabilistic arguments, as one realizes a posteriori analysing the final result Eq. \eqref{eq:Fsc_finalFormula} obtained with the replica formalism. In fact, as already discussed at the end of the previous Section, at the $O(1/N)$ order  loops are sparsely distributed in the graph and do not interact with each other. Therefore their contributions to the free energy  can be summed up separately and each one of them can be considered as  embedded in an infinite tree. In order to compute the free energy shift due to the presence of a loop of length $\ell$, we consider a very large random tree, with partition function $Z_{T}$, and remove the $\ell+1$ edges of an open chain of length $\ell+1$, as showed in Figure \ref{fig:cavity}. We call  $\s_0,\dots,\s_\ell+1$ the cavity spins of the new graph, that is the ones who lost one (this is the case of $\s_0$ and $\s_{\ell+1}$) or two ($\s_1,\dots,\s_{\ell}$) of their adjacent edges. We call $Z_{cav}(\s_0,\dots,\s_{\ell+1})$ the partition function of this new system, conditioned on the values of the cavity spins. Since we assumed to start from a tree graph, the partition function $Z_{cav}$ takes the form
\begin{equation}
Z_{cav}(\s_0,\dots,\s_{\ell+1})= \tilde Z e^{h_0\s_0+r_1\s_1+\ldots+r_\ell\s_\ell+h_{\ell+1}\s_{\ell+1}},
\end{equation}
where $\tilde Z\geq 0$ and the cavity fields  $h_i$ and $r_{0/\ell+1}$ are independently distributed according to $P(h)$ from \eqref{eq:cavityEq} and $R(r)$ from Eq. \eqref{eq:cavityEq_for_r} respectively. We recover the partition function of the original tree adding back the missing links, therefore we establish the relation
\begin{equation}
Z_{T}= \tilde Z\times Z^o_{\ell+1},
\end{equation}
where $Z^o_{\ell+1}$ is the partition function of an open chain of length $\ell+1$ with incoming fields $h_0,r_1,\dots,r_{\ell},h_{\ell+1}$. On the other hand, starting from the cavity graph, we can create another graph $G$ containing exactly one loop. This can be achieved adding an edge between the spin $\s_0$ and $\s_{\ell+1}$, and adding other $\ell$ edges to form a loop among the $\ell$ internal cavity spins (see Figure \ref{fig:cavity}). Notice that with this construction all the spins retain the same degree that they had in the original graph $T$. The partition function of the system defined on $G$ is then  given by 
\begin{equation}
Z_{G}= A\times Z^o_{1}\times Z^c_{\ell}.
\end{equation}
We are interested in the difference of the average free energy between the system $G$ an $T$ in the large graph limit. Let us call $\mathrm N$ the number of nodes in $T$ and $G$. The free energy shift is then given by
\begin{equation}
\Delta \phi_\ell = -\frac{1}{\beta}\,\lim_{\mathrm N\to\infty}\,[\log Z_G - \log Z_T]_{\mathrm{av}}.
\label{phi1}
\end{equation}
For the average free energy $\phi^o_L$ of an open chain of length $L$ embedded in a RRG  the following relation holds\cite{Lucibello2014}:
\begin{equation}
\phi^o_L =  L\, \phi + \phi_{s},
\end{equation}
where $\phi_s$  is a site term that does not depend on $L$ \cite{Lucibello2014}. It is therefore easy to derive the expected result:
\begin{equation}
\Delta \phi_\ell= \phi^c_\ell - \ell\, \phi.
\label{phi2}
\end{equation}
We have proven that the free energy difference $\Delta \phi_\ell$ as defined by Eq. \eqref{phi1}  corresponds to the quantity $\phi^c_\ell - \ell \phi$, as it was defined in the last Section.
Taking into account that the average number of loops of length $\ell$ in a graph of the RRG ensemble is $\frac{z^\ell}{2\ell}$ in the thermodynamic limit, we re-obtain Eq. \eqref{eq:Fsc_finalFormula} without making any resort to replicas. 

The argument we gave in this Section to compute the first finite size correction to the free energy is strictly limited to the RRG ensemble. In fact it relies heavily on the homogeneity of the graphs. On different graph ensembles more refined combinatorial arguments, as the one given in \cite{Ferrari2013} for Erd\"{o}s-R\'enyi random graphs, have to be used.

\section{Numerical Experiment: Spin Glass in a Magnetic Field}
\label{sec:num}

\begin{figure*}
\includegraphics[width=\textwidth]{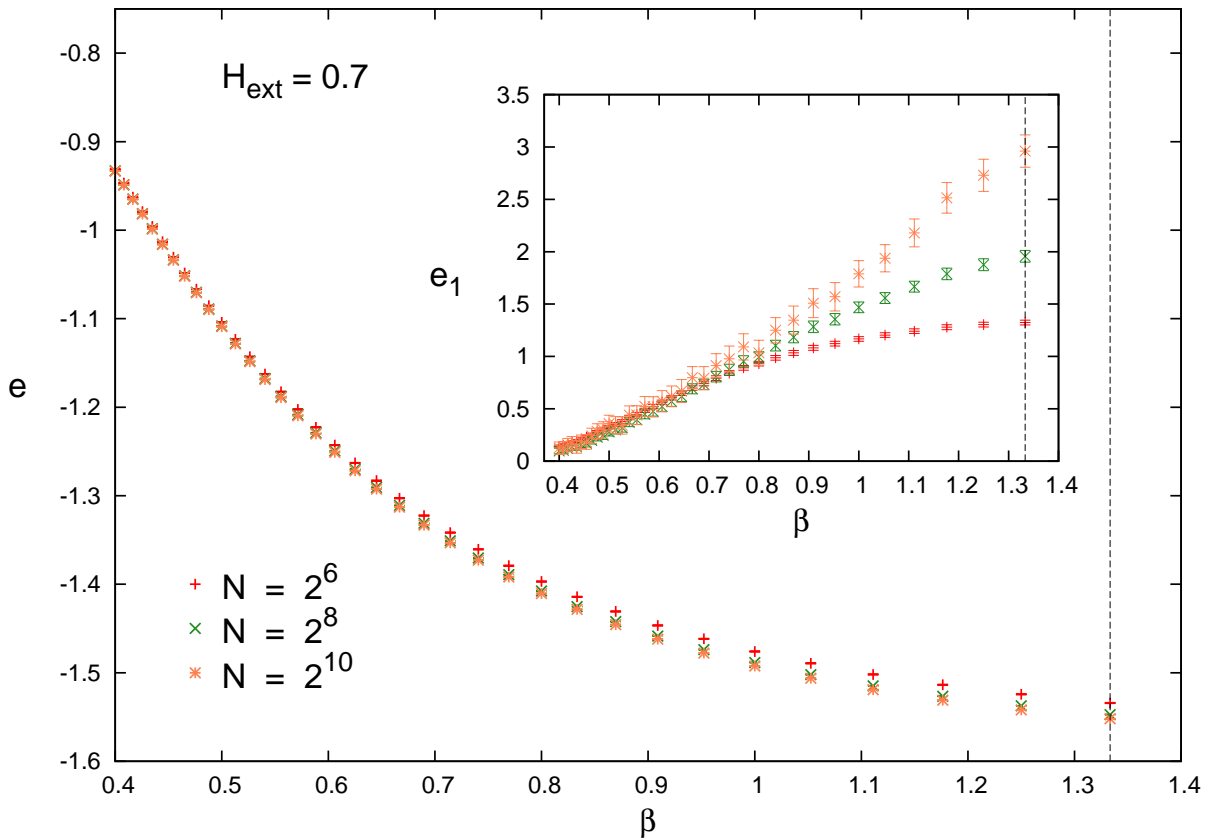}
\caption{Main panel: average energy density $e(\beta)$ of a spin glass on a RRG with connectivity $c=4$,  bimodal random couplings $J=\pm1$ and uniform external magnetic field $H_{ext} = 0.7$.
 The profiles of $e(\beta)$ are drawn for different system sizes: from top to bottom $N=64, 256, 1024$. Inset: the coefficient $e_1(\beta)$ of the $1/N$ correction to the thermodynamic energy density, measured experimentally by means of Eq. \eqref{eq:diff}. 
The vertical dashed lines, both in the main panel and in the inset, mark the position of the  critical point. Notice that the coefficient $e_1$ does not depend on the size $N$ only up to $\beta \sim 0.8$. Above this value there is an evident $N$-dependence, which is more and more pronounced as one gets closer to the critical point. Exactly at the critical temperature, the coefficient $e_1$ diverges when the system size $N$ goes to infinity. This divergence is the signal of the onset of a different type of scaling of the finite size corrections at the critical point and, indeed, in all the critical domain.
\label{fig:leading_order}}
\end{figure*}

\begin{figure*}
\includegraphics[width=\textwidth]{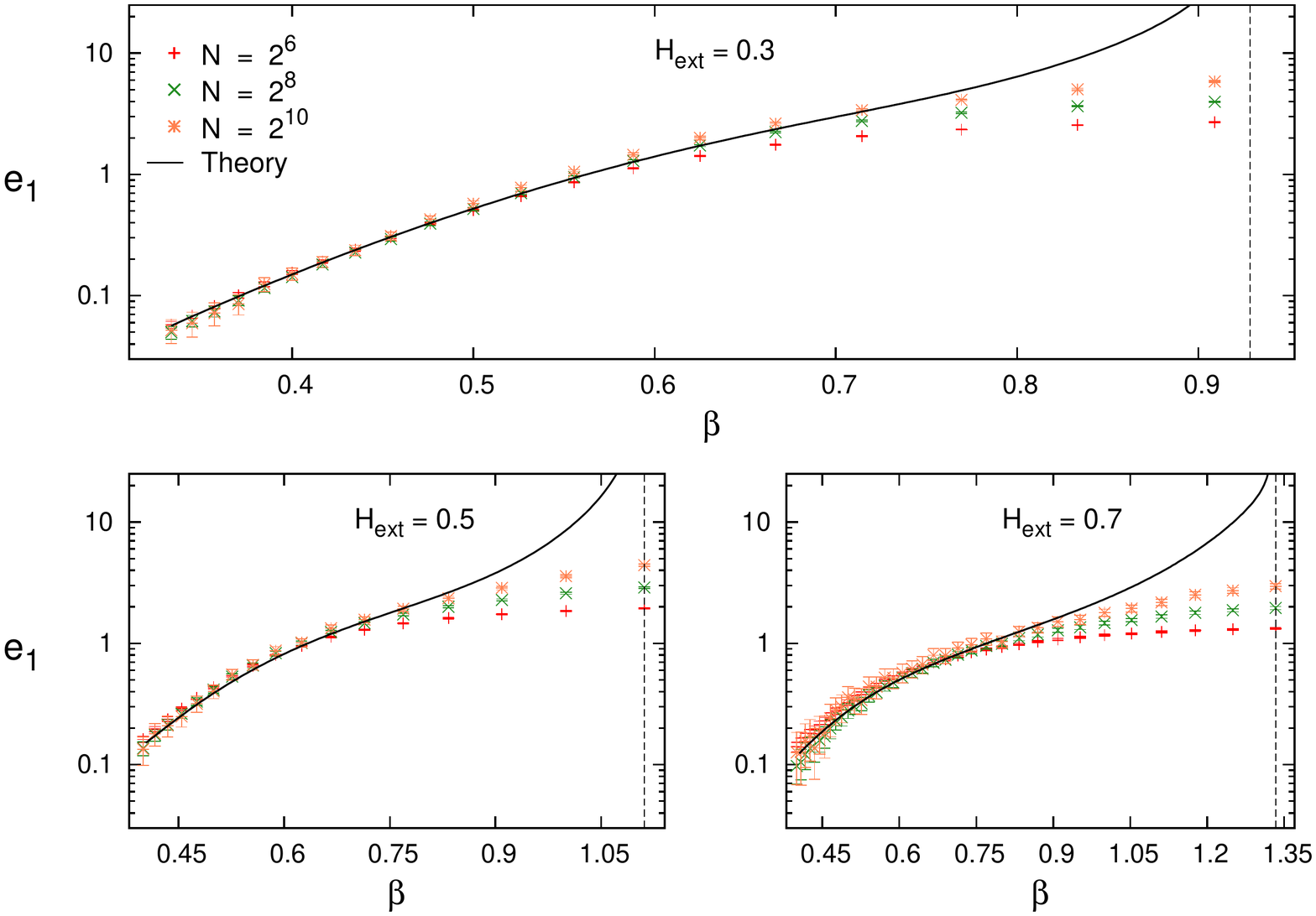}
\caption{ Finite size corrections to the energy density of a spin glass model on a RRG with connectivity $c=4$ and bimodal random couplings $J=\pm1$. The various panels refer to different values of the external uniform magnetic field. The results obtained from Monte Carlo simulations are compared with the analytical values predicted from Eqs. \eqref{eq:Fsc_finalFormula} and \eqref{e1}. The vertical dashed lines mark the positions of the  critical temperatures.
\label{fig:correzioni}}
\end{figure*}

In this Section we test our analytical prediction for the finite size correction to the free energy, Eq. \eqref{eq:Fsc_finalFormula}, on the spin glass in a uniform magnetic field. The connectivity of the graph is $c=4$. In the experiment the couplings $J_{ij}$ are bimodal random variables, taking value $J_{ij}=\pm1$ with equal probability. We simulate the model using a parallel tempering Monte Carlo algorithm and three different values of the external field $H=0.3,0.5$ and $0.7$. For each value of the magnetic field $H$ we simulate systems of three different sizes: $N=2^6,2^8$ and $2^{10}$. The numerical estimate of the coefficient $f_1$ of the $O(1/N)$ correction is obtained as the difference between the free energies of systems of different system sizes, viz.:
\be
2N[f(N)\ -\ f(2N)]\ = \ f_1\ +\ o(1)\ .
\label{eq:diff}
\ee
The $o(1)$ term in the r.h.s. of Eq.\eqref{eq:diff} accounts for subleading corrections. These subdominant contributions become particularly important at the critical point, but also in all the critical domain (see \figurename\ref{fig:correzioni}). The situation is more involved below the critical point, where the leading corrections have a totally different scaling (no more proportional to $1/N$) and, consequently, our theoretical prediction does not hold anymore.

In order to compute the analytical estimate of $ f_1$ we proceed in two steps: we explicitly calculate the first terms of the sum. We computed by transfer matrix multiplication the partition function and the free energy of closed chain of length $\ell$, for many realizations of the disorder and up to $\ell=7$. We then resummed the remaining terms of the series using the criterion explained in Ref. \cite{Ferrari2013}, that we briefly recap.
Using the formalism of the replicated transfer matrix developed in Ref.\cite{Lucibello2014}, one can show that, in a spin glass, the dominant contribution to $ f_1$ comes from the replicon eigenvalue. Therefore, we use only the knowledge of this eigenvalue to analytically resum the remaining terms of the series (from $\ell=8$ to $\infty$).  The large $\ell$ behaviour of the shift $\Delta\phi_\ell$ is given by the  expression
\be
\Delta\phi_\ell \sim A \lambda^{\ell} \qquad  \mathrm{for}\quad   \ell\gg1 ,
\ee
where $\lambda$ is the replicon eigenvalue,  the largest eigenvalue satisfying the following integral equation:
\be
\lambda g_{\lambda}(u) = \mathbb{E}_{J,r}\int\dd u'\ g_{\lambda}(u') \delta[u-\hat{u}(\beta J,r+u')] 
\left(
\frac{\partial\hat{u}}{\partial u}
\right)^2.
\ee
Here $r$ is distributed as $R(r)$ defined in Eq. \eqref{eq:cavityEq_for_r}. The maximum eigenvalue of the integral operator in last equation can be obtained numerically by population dynamics techniques.
The coefficient $A$ instead can be computed analytically, as shown in Ref.\cite{Lucibello2014}, and takes value $A=3/(2\beta)$.
We can split the quantity $ f_1$ in two pieces:
\be
 f_1 \sim  S(L) - \frac{3}{4\beta} \mathcal{L}\textrm{og}_{L+1}\left[
1-(c-1)\lambda
\right] ,
\ee
where $S(L)$ is the partial sum over the loops up to $\ell=L$, and the second term is the resummation of the remaining series from $\ell=L+1$ to $\ell=\infty$, which we represented via the function $\mathcal{L}\textrm{og}_{p}(1-x)$, defined as:
\be
\mathcal{L}\textrm{og}_{p}(1-x) = -\sum_{\ell=p}^{\infty}\frac{x^\ell}{\ell} .
\ee
In our concrete case we can compute explicitly the first $L=7$ terms of the series, and so, the approximated analytic form of $ f_1$ is 
\be
f_1 \sim  S(7) - \frac{3}{4\beta} \mathcal{L}\textrm{og}_{8}\left[
1-3\lambda
\right] \qquad   \mathrm{for}\quad   c=4 .
\ee

In a numerical simulation, measuring the energy is, actually, much simpler than the free energy (since the last one involves an estimate of the entropy). As a consequence we preferred to compare analytical and numerical results for the finite size corrections to the energy density $ e_1$. Analytically , the quantity $ e_1$ is given by the usual formula relating energy and free energy:
\be
e_1 = f_1+\beta \frac{\partial  f_1}{\partial\beta}.
\label{e1}
\ee

In Figure \ref{fig:correzioni} we show the comparison between the experiments and our theoretical result. The agreement is good at high temperatures, while it deteriorates close to the critical point. At the critical point in fact every order of the $O(1/N)$ expansion of the free energy diverges, therefore near the critical point subleading finite size corrections become increasingly important and extrapolation of $ e_1$ obtained from numerical simulations to its large $N$ limit, that can be derived by our analytical expression \eqref{eq:Fsc_finalFormula}, is difficult to achieve.

Below the critical point, the nature of the finite size corrections changes dramatically, because the replica symmetry is broken in the spin glass phase and it is widely believed that the correct solution of the model is obtained by using the Parisi hierarchical breaking pattern \cite{Parisi1987} in the same way it is used to solve the fully connected version of the model. The solution in the spin glass phase has the property to be marginally stable and the finite size corrections can be assessed by computing the volume of zero modes. 
This should imply a finite size correction to the free energy density proportional to $N^{-2/3}$ below the critical point, instead of the simple $1/N$ found in the paramagnetic phase.
Another possible source of finite size corrections, of the same magnitude of the previous one, could be the existence of other solutions to the saddle point equations, which are characterized by a different replica symmetry breaking pattern. When resummed, these solutions, although thermodynamically irrelevant, can give finite size effects comparable to those generated by the integration over the Goldstone modes.

\section{Conclusions}
In this work we derived an analytical expression for the $O(1/N)$ correction to the average free energy of Ising disordered systems on random regular graphs. This correction, Eq. \eqref{eq:Fsc_finalFormula}, is expressed as a weighted sum over the loops of the graph. Each loops contributes according to the free energy shift due to its  addition to an infinite tree, as showed in  Eq. \eqref{phi1}). We compared our analytical predictions with numerical simulations on a spin glass model and obtained an excellent agreement in the region where sub-leading finite size effects are small. 

We argue that the form of the  $O(1/N)$ finite size corrections, given in Eq. \eqref{eq:Fsc_finalFormula}, is independent of the specific structure of the model (namely Ising spins), as other recent works also confirm\cite{Ferrari2013}\cite{Metz2014}, but depends only on its topological features. 

 Moreover it is possible to extend the formalism we presented to produce a perturbative expansion, around the Bethe free energy, for disordered systems  on finite dimensional lattices. This is currently being investigated by the authors, applying the replica method to a scheme resembling the one proposed in Ref. \cite{Vontobel2013}.

The research leading to these results has received funding from the European Research Council under the European Union's Seventh Framework Programme (FP7/2007-2013) / ERC grant agreement No. 247328 and from the
Italian Research Minister through the FIRB project No. RBFR086NN1.

\bibliography{bibliography}

\appendix

\begin{widetext}
\section{Integral representation of $[Z^n]_{\mathrm{av}}$}
\label{app:App1}

The average of the replicated partition function of the model reads:
\be
[Z^n]_{\mathrm{av}}=
\left[
\sum_{\{\s\}}\left(
\prod_{i<j}\exp\left(\beta J_{ij}C_{ij}\sum_{a=1}^{n}\s_i^a\s_j^a\right)
\prod_i\exp\left(\beta H_i\sum_{a=1}^{n}\s_i^a\right)
\right)
\right]_{\mathrm{av}},
\label{eq:Zn1}
\ee
where the sum has to be taken over the  replicated spins $\{\s_i^a\}$ for $i=1,\dots,N$ and $a=1,\dots,n$. The average $[\ \cdot\ ]_{\mathrm{av}}$ has to be performed over the RRG ensemble, the couplings $J_{ij}$ and the random fields $H_i$. The graph ensemble is defined by the probability of sampling one of its element, having adjacency matrix $C_{ij}$. This is given by
\be
\mathcal{P}(C)=\frac{1}{\mathcal{N}}\prod_{i<j}
\left[
\left(1-\frac{c}{N}
\right)
\delta(C_{ij})+
\frac{c}{N}\delta(C_{ij}-1)
\right]
\prod_i
\delta\left(\sum_{j\neq i}C_{ij}-c\right).
\label{eq:P(C)}
\ee
Here the variable $c$ is the connectivity of the nodes, the weighting factors $1-\frac{c}{N}$ and $\frac{c}{N}$ have been chosen for convenience and $\mathcal{N}$ is a normalization factor. Let us define
\be
\begin{aligned}
U(\s_i,\s_j)&=\mathbb{E}_J
\left(
e^{\beta J\sum_a\s_i^a\s_j^a}
\right), \\
B(\s_i)&=\log\mathbb{E}_H
\left(
e^{\beta H\sum_a\s_i^a}
\right).
\end{aligned}
\ee
The arguments of the functions $U(\s_i,\s_j)$ and $B(\s_i)$ indicate the replicated spins $\s_i\equiv(\s_i^1,\dots,\s_i^n)$. When it can cause confusion, the replica label $a$ will be explicitly written.

After averaging over the graph ensemble \cite{Dean2000}, Eq. \eqref{eq:Zn1} takes the following form:
\be
[Z^n]_{\mathrm{av}}=\frac{1}{\mathcal{N}}\sum_{\{\s\}}
\int\left(\prod_i\dd\lambda_i\ e^{-i\lambda_ic+B(\s_i)}\right)
\exp\left\{
\sum_{i<j}\log\left[
1+\frac{c}{N}\left(
e^{i\lambda_i}U(\s_i,\s_j)e^{i\lambda_j}-1
\right)
\right]
\right\},
\label{eq:Zn2}
\ee
where we used the integral representations of the Kronecker $\delta$-functions appearing in $\mathcal{P}(C)$:
\be
\prod_i
\delta\left(\sum_jC_{ij}-c\right)=\int_0^{2\pi}\prod_i\frac{\dd\lambda_i}{2\pi}\ e^{i\lambda_i\left(\sum_{j\neq i}C_{ij}-c\right)}.
\ee

We expand the argument of the exponential in Eq. \eqref{eq:Zn2} to obtain
\be
\begin{aligned}
\sum_{i<j}\log\left[
1+\frac{c}{N}\left(
e^{i\lambda_i}U(\s_i,\s_j)e^{i\lambda_j}-1
\right)
\right]&=
\left(\frac{c}{2N}+\frac{c^2}{2N^2}\right)\left[\sum_{ij}e^{i\lambda_i}U(\s_i,\s_j)e^{i\lambda_j}\right]-\frac{c}{2N}\left[\sum_{i}e^{2i\lambda_i}U(\s_i,\s_i)\right]\\
&-\frac{c^2}{4N^2}
\left[\sum_{ij}e^{2i\lambda_i}U^2(\s_i,\s_j)e^{2i\lambda_j}\right]+A(N,c)+ O\left(\frac{1}{N}\right),
\end{aligned}
\label{eq:LogExpansion}
\ee
where the constant $A(N,c)$ is given by
\be
A(N,c)=-\frac{cN}{2}+\frac{c}{2}-\frac{c^2}{4}.
\ee 

In order to compute the sum over the spin variables in Eq. \eqref{eq:Zn2}, we need to decouple the sites. Site factorization can be achieved by introducing two functions $\rho_1(\s)$ and $\rho_2(\s)$, defined as 
\be
\begin{aligned}
\rho_1(\s)&=\frac{1}{N}\sum_i e^{i\lambda_i}\prod_a\delta(\s^a-\s_i^a) ,\\
\rho_2(\s) &= \frac{1}{N}\sum_i e^{2i\lambda_i}\prod_a\delta(\s^a-\s_i^a),
\end{aligned}
\ee 
if we are interest only in the first correction in $\frac{1}{N}$. To obtain higher orders in the expansion we would need up to $c$ different functions $\rho_k(\s)$, as will be clear from what follows.
Using $\rho_1(\s)$ and $\rho_2(\s)$, along with the expansion \eqref{eq:LogExpansion}, Eq.  \eqref{eq:Zn2} becomes
\begin{align}
[Z^n]_{\mathrm{av}} \sim \sum_{\{\s\}}\exp&\left\{
\frac{Nc}{2}\left(1+\frac{c}{N}\right)\int\dd\s\,\dd\tau\ \rho_1(\s)U(\s,\tau)\rho_1(\tau) + 
N\log\int\frac{\dd\lambda}{2\pi}\,\dd\s\ \exp\left[
-ic\lambda+B(\s)-\frac{c}{2N}e^{2i\lambda}U(\s,\s)
\right]\right. \nonumber\\
&\left. - \frac{c^2}{4}\int\dd\s\,\dd\tau\ \rho_2(\s)U^2(\s,\tau)\rho_2(\tau) + A(N,c) - \log\mathcal{N}
\right\},
\label{eq:Zn3}
\end{align}
up to the order $O(1)$.
In the previous equation the notation $\int\dd\s$ stands for:
\be
\int\dd\s \equiv \prod_{a=1}^{n}\sum_{\s^a=\pm1} .
\ee
We note also that the function $U(\s,\s)$, evaluated on the same first and second arguments, is independent on $\s$. To lighten the notation we then define:
\be
U(\s,\s) = \mathbb{E}_J e^{n\beta J} \equiv U_0 .
\ee

Proceeding in the calculation, we introduce two $\delta$-functionals to enforce the definitions of $\rho_1(\s)$ and $\rho_2(\s)$:
\be
1 = \int\mathcal{D}\rho_k\ \delta\left[
\rho_k(\s)-\frac{1}{N}\sum_ie^{ik\lambda_i}\prod_a\delta(\s^a-\s_i^a)
\right]   \qquad\mathrm{for }\quad  k=1,2 .
\ee
The functional measure $\mathcal{D}\rho$ is defined as $\mathcal{D}\rho\equiv\prod_{\s\in \mathbb{R}^n}\dd\rho(\s)$. Moreover, we use the following integral representation of the $\delta$-functional:
\be
\delta[\rho] = \int\mathcal{D}\hat{\rho}\ e^{-\int\dd\s\ \rho(\s)\hat{\rho}(\s)} ,
\ee
where $\mathcal{D}\hat\rho\equiv\prod_{\s\in \mathbb{R}^n}\frac{\dd\hat\rho(\s)}{2\pi}$, to rewrite the replicated partition function \eqref{eq:Zn3} as
\begin{align}
[Z^n]_{\mathrm{av}} \sim &\int\left(\prod_{k=1}^2\mathcal{D}\rho_k\mathcal{D}\hat{\rho}_k\right) 
\exp\Bigg\{
-N\int\dd\s\ \rho_1(\s)\hat{\rho}_1(\s)+\frac{Nc}{2}\left(1+\frac{c}{N}\right)
\int\dd\s\,\dd\tau\ \rho_1(\s)U(\s,\tau)\rho_1(\tau)-\int\dd\s\ \rho_2(\s)\hat{\rho}_2(\s)\nonumber
\\
& - 
\frac{c^2}{4}\int\dd\s\,\dd\tau\ \rho_2(\s)U^2(\s,\tau)\rho_2(\tau) + 
N\log\int\dd\s\ e^{B(\s)}\int\frac{\dd\lambda}{2\pi}\ \exp\left[
\hat{\rho}_1(\s)e^{i\lambda}-ic\lambda+\frac{e^{2i\lambda}}{N}\left(
\hat{\rho}_2(\s)-\frac{cU_0}{2}
\right)
\right]\nonumber\\
& + A(N,c)-\log\mathcal{N}
\Bigg\} .
\label{eq:Zn4}
\end{align}
We now can carry out the $\lambda$ integration, expanding the exponential and obtaining
\be
\int_0^{2\pi}\frac{\dd\lambda}{2\pi}\ \exp\left[
\hat{\rho}_1(\s)e^{i\lambda}-ic\lambda+\frac{e^{2i\lambda}}{N}\left(
\hat{\rho}_2(\s)-\frac{cU_0}{2}
\right)
\right] = \sum_{m=0}^{c/2}\frac{1}{m!(c-2m)!}\frac{\hat{\rho}_1(\s)^{c-2m}}{N^m}
\left(
\hat{\rho}_2(\s)-\frac{cU_0}{2}
\right)^m .
\label{eq:lambdaInt}
\ee
In the sum on the r.h.s. of Eq. \eqref{eq:lambdaInt} we retain only the leading  terms, corresponding to $m=0$ and $m=1$. The partition function \eqref{eq:Zn4} now reads
\begin{align}
[Z^n]_{\mathrm{av}} \sim &\int\left(\prod_{k=1}^2\mathcal{D}\rho_k\mathcal{D}\hat{\rho}_k\right) 
\exp\Bigg\{
-N\int\dd\s\ \rho_1(\s)\hat{\rho}_1(\s)+\frac{Nc}{2}\left(1+\frac{c}{N}\right)
\int\dd\s\,\dd\tau\ \rho_1(\s)U(\s,\tau)\rho_1(\tau)-\int\dd\s\  \rho_2(\s)\hat{\rho}_2(\s)\nonumber
\\
& - 
\frac{c^2}{4}\int\dd\s\,\dd\tau\ \rho_2(\s)U^2(\s,\tau)\rho_2(\tau) + 
N\log\int\dd\s\ e^{B(\s)}\hat{\rho}_1(\s)^c + c(c-1)\frac{\int\dd\s e^{B(\s)}\hat{\rho}_1(\s)^{c-2}
\left(\hat{\rho}_2(\s)-\frac{cU_0}{2}\right)}
{\int\dd\s\ e^{B(\s)}\hat{\rho}_1(\s)^{c}}
\nonumber\\
& + A(N,c) - \log\mathcal{N} - N\log c!
\Bigg\} .
\end{align}
At this point it is natural to define a new field $r(\s)$ as
\be
r(\s) = c^2 \frac{\int\dd\s\ e^{B(\s)} \hat{\rho}_1(\s)^{c-2}
}
{\int\dd\s\ e^{B(\s)}\hat{\rho}_1(\s)^{c}} ,
\ee
and to observe that the integral over $\hat{\rho}_2(\s)$ gives
\be
\int\mathcal{D}\hat{\rho}_2\ \exp\left\{
-\int\dd\s \hat{\rho}_2(\s)\left[
\rho_2(\s)-\frac{c-1}{c} r(\s)
\right]
\right\} = 
\delta\left[
\rho_2-\frac{c-1}{c} r
\right] .
\ee
Integrating out also $\rho_2$, we obtain
\begin{align}
[Z^n]_{\mathrm{av}} \sim &\int\mathcal{D}\rho_1\mathcal{D}\hat{\rho}_1 
\exp\Bigg\{
-N\int\dd\s\ \rho_1(\s)\hat{\rho}_1(\s)+\frac{Nc}{2}\left(1+\frac{c}{N}\right)
\int\dd\s\,\dd\tau\ \rho_1(\s)U(\s,\tau)\rho_1(\tau)\nonumber
\\
& + 
N\log\int\dd\s\ e^{B(\s)}\hat{\rho}_1(\s)^c 
- \frac{(c-1)^2}{4}\int\dd\s\,\dd\tau\ r(\s)U^2(\s,\tau)r(\tau) 
- \frac{(c-1)U_0}{2}\int\dd\s\ r(\s)
\nonumber\\
& + A(N,c) - \log\mathcal{N} - N\log c!
\Bigg\} .
\end{align}

The integral over $\rho_1$ is Gaussian and can be performed explicitly and we get:
\begin{align}
[Z^n]_{\mathrm{av}} \sim 
[\det(cU)]^{-1/2}&\int\mathcal{D}\hat{\rho}_1 
\exp\Bigg\{
-\frac{Nc}{2}
\int\dd\s\,\dd\tau\ \hat{\rho}_1(\s)U^{-1}(\s,\tau)\hat{\rho}_1(\tau) + 
N\log\int\dd\s\ e^{B(\s)}\hat{\rho}_1(\s)^c \nonumber\\
&+\frac{1}{2}\int\dd\s\,\dd\tau\ \hat{\rho}_1(\s)U^{-1}(\s,\tau)\hat{\rho}_1(\tau) - 
 \frac{(c-1)^2}{4}\int\dd\s\,\dd\tau\ r(\s)U^2(\s,\tau)r(\tau)
\nonumber\\
 &- \frac{(c-1)U_0}{2}\int\dd\s\ r(\s) + A(N,c) - \log\mathcal{N} - N\log c!
\Bigg\} .
\end{align}
We make the following change of variables, introducing at last the order parameter  $\rho(\s)$ through
\be
\hat{\rho}_1(\s) = c\int\dd\s\ U(\s,\tau)\rho(\tau) ,
\ee
which redefines also the field $r(\s)$ as
\be
r(\s) = \frac{
e^{B(\s)}\left[
\int\dd\tau\ U(\s,\tau)\rho(\tau)
\right]^{c-2}
}
{
\int\dd\s\ e^{B(\s)}\left[
\int\dd\tau\ U(\s,\tau)\rho(\tau)
\right]^{c}
} .
\label{eq:Equzione_per_il_campo_r}
\ee

The computation of the factor $\log\mathcal{N}$ can be done along the same lines of the preceding derivation, and gives \cite{Dean2000}:
\be
\log\mathcal{N} \sim N(c\log c-\log c!-c) + \frac{c}{2} + \frac{1}{4} - \frac{\log 2}{2}.
\ee
Calling for brevity $\mathcal{A}(N,c)$ the  quantity
\be
\begin{aligned}
\mathcal{A}(N,c) &= A(N,c) + Nc\log c - N\log c! - \log\mathcal{N} \\
&= \frac{Nc}{2} - 
\frac{c^2+1}{4} + \frac{\log2}{2} ,
\end{aligned}
\ee 
we obtain the final expression of the $[Z^n]_{\mathrm{av}}$ up to the order $O(1)$, that is
\be
[Z^n]_{\mathrm{av}} \sim [\det(cU)]^{1/2} e^{\mathcal{A}(N,c)} 
\int\mathcal{D}\rho\ e^{-N\mathcal{S}_0[\rho] - \mathcal{S}_1[\rho]}.
\ee
Here the integration measure is given by $\mathcal{D}\hat\rho\equiv\prod_{\s\in \mathbb{R}^n}\frac{\dd\rho(\s)}{\sqrt{2\pi}}$, and the  functionals $\mathcal{S}_0[\rho]$ and $\mathcal{S}_1[\rho]$ read
\be
\begin{aligned}
\mathcal{S}_0[\rho] &= \frac{c}{2}\int\dd\s\,\dd\tau\, \rho(\s)U(\s,\tau)\rho(\tau) - 
\log\int\dd\s\ e^{B(\s)}\left[
\int\dd\tau\ U(\s,\tau)\rho(\tau)
\right]^c ,\\
\mathcal{S}_1[\rho] &= -\frac{c^2}{2}\int\dd\s\,\dd\tau\ \rho(\s)U(\s,\tau)\rho(\tau) + 
\frac{(c-1)^2}{4}\int\dd\s\,\dd\tau\ r(\s)U^2(\s,\tau)r(\tau) + \frac{(c-1)U_0}{2}\int\dd\s\ r(\s) .
\end{aligned}
\ee

\section{Computing the finite size corrections}
\label{app:App2}

There are two sources for the $1/N$ finite size corrections to the thermodynamic free energy density. The first contribution comes from the subleading part of the replicated action $\mathcal{S}_1[\rho]$, evaluated in the saddle point solution $\rho_*$, given in eq. \eqref{eq:SPequation}.
The second one stems from the Gaussian integral obtained by expanding the leading action $\mathcal{S}_0[\rho]$ around the saddle point. This is given by
\begin{equation}
 [\det(cU)]^{1/2} 
\int\mathcal{D}\chi\ e^{-\frac{1}{2}\int\chi\,\partial^2\mathcal{S}^*_0\,\chi} =e^{-\frac{1}{2}\log\det(\mathbb{I}-\Sigma)},
\end{equation}
where $\chi(\s)$ is the rescaled fluctuation of $\rho(\s)$ around the saddle point and we omitted the dependence on the replicated spins in the exponent of the l.h.s.. The matrix $\Sigma(\s,\tau)$ is defined as
\be
\begin{aligned}
&\Sigma(\s,\tau) = (c-1)T(\s,\tau)-c\left[
\int\dd\s'\ U(\s,\s')\rho_*(\s')
\right]\rho_*(\tau) ,\\
&T(\s,\tau) = U(\s,\tau)r_*(\tau) .
\end{aligned}
\ee
 Summing up all the contributions, the averaged replicated partition function $[Z^n]_{\mathrm{av}}$ up to order $O(1)$ becomes:
\be
[Z^n]_{\mathrm{av}} \sim \exp\left[\mathcal{A}(N,c)
-N\mathcal{S}_0[\rho_*]-\mathcal{S}_1[\rho_*]+
\frac{1}{2}\sum_{\ell=1}^{\infty}\frac{1}{\ell}\tr \Sigma^\ell
\right].
\label{eq:ZrepAllordine1suN}
\ee 

The introduction of the auxiliary matrix $T(\s,\tau)$ will simplify a lot the computation of the trace appearing in Eq. \eqref{eq:ZrepAllordine1suN}. The crux is to observe that $\rho_*(\s)$ is a left eigenvector of $T(\s,\tau)$ with eigenvalue $1$, i.e.
\be
\int\dd\s\ \rho_*(\s) T(\s,\tau) = \rho_*(\tau).
\ee
This property can be verified by acting on the left with $T(\s,\tau)$ on the saddle point equation \eqref{eq:SPequation}.

It is useful to define also an auxiliary function $\hat{\rho}_*(\s)$ as follows:
\be
\hat{\rho}_*(\s) = \int\dd\tau\ U(\s,\tau)\rho_*(\tau) .
\ee
As a consequence of the saddle point equation \eqref{eq:SPequation}, the function $\hat{\rho}_*(\s)$ has the following interesting property:
\be
\int\dd\s\ \hat{\rho}_*(\s) \rho_*(\s) = 1 .
\ee
Using the definition of $T(\s,\tau)$ and $\hat{\rho}_*(\s)$, the matrix $\Sigma(\s,\tau)$ can be cast in a simpler form, that reads
\be
\Sigma(\s,\tau) = (c-1)T(\s,\tau) - c \hat{\rho}_*(\s) \rho_*(\tau) .
\label{eq:SigmaSimple}
\ee
The two matrices  $T(\s,\tau)$ and $\hat{\rho}_*(\s) \rho_*(\tau)$ in the r.h.s of Eq. \eqref{eq:SigmaSimple} they do commute with each other, as can be checked by inspection. Therefore, the trace of the $\ell$-th power of the matrix $\Sigma$ can be written as 
\be
\tr \Sigma^\ell =
\sum_{k=0}^{\ell}\binom{\ell}{k}(c-1)^{\ell-k}(-c)^k \tr\left[
\left(
\hat{\rho}_*\rho_*
\right)^k
T^{\ell-k}
\right] .
\label{eq:TracciaSigma}
\ee
Observing that in all the terms of the sum, but the one corresponding to $k=0$, the matrix $T$ is multiplied on the left by its left eigenvector with unitary eigenvalue, we  easily get the following result:
\be
\tr\left[
\left(
\hat{\rho}_*\rho_*
\right)^k
T^{\ell-k}
\right] = \begin{cases}
\tr T^\ell &\textit{for  } k=0\\
   1 & \textit{for  } k\neq0 .
\end{cases}
\ee
We can now immediately evaluate Eq. \eqref{eq:TracciaSigma} and we find:
\be
\tr \Sigma^\ell = (c-1)^\ell
\left[
\tr T^\ell-1
\right]+
(-1)^\ell .
\label{eq:tracciaSigma}
\ee
Inserting Eq. \eqref{eq:tracciaSigma} into Eq. \eqref{eq:ZrepAllordine1suN} we get
\be
[Z^n]_{\mathrm{av}} \sim \exp\left[
-N\mathcal{S}_0[\rho_*]-\mathcal{S}_1[\rho_*]+
\frac{1}{2}\sum_{\ell=1}^{\infty}\frac{(c-1)^\ell}{\ell}
\left[
\tr T^\ell -1
\right] + \mathcal{A}(N,c)-\frac{\log2}{2}
\right] .
\label{eq:ZrepAllordine1suN_2}
\ee
The term $\mathcal{S}_1[\rho_*]$ can be expressed using the matrix $T$ in the following way:
\be
\mathcal{S}_1[\rho_*] = \frac{c^2+1}{4} + \frac{(c-1)^2}{4}\left[\tr T^2-1\right] + 
\frac{c-1}{2}\left[\tr T-1\right] .
\ee
Therefore we see that the terms $\ell=1$ and $\ell=2$ in the sum in Eq. \eqref{eq:ZrepAllordine1suN_2} cancel out with the terms coming from $\mathcal{S}_1[\rho_*]$. Moreover, using the definiton of $\mathcal{A}(N,c)$, and the noting that $\mathcal{S}_0[\rho_*]$ equals
\be
\mathcal{S}_0[\rho_*] = \frac{c}{2} - \log\int\dd\s\ e^{B(\s)}\left[
\int\dd\tau\ U(\s,\tau)\rho_*(\tau)
\right]^c ,
\ee
we finally obtain
\be
[Z^n]_{\mathrm{av}} \sim \exp\left\{
N\log\int\dd\s\ e^{B(\s)}\left[
\int\dd\tau\ U(\s,\tau)\rho_*(\tau)
\right]^c + 
\frac{1}{2}\sum_{\ell=3}^{\infty}\frac{(c-1)^\ell}{\ell}
\left[
\tr T^\ell-1
\right]\
\right\} .
\label{eq:ZrepAllordine1suN_3}
\ee

We split the free energy density $f(N)$ into the sum of the leading term plus the $1/N$ correction:
\be
f(N) = f_0 + \frac{1}{N} f_1 + o(1/N) .
\ee

The quantity $f_1$ is given by
\be
f_1 = -\frac{1}{\beta}\lim_{n\to0} 
\sum_{\ell=3}^{\infty}\frac{(c-1)^\ell}{2\ell}
\,\partial_n
\tr T^\ell.
\label{eq:DeltaFprimaDelLimite}
\ee

\section{Evaluating $\tr T^\ell$}
\label{app:tr}
The matrix $T(\s,\tau)$ is defined as $T(\s,\tau)=U(\s,\tau)r(\tau)$. In the replica-symmetric regime, we can parametrize the field $r(\s)$ as
\be
r(\s)=\int\dd r\ R_n(r) \frac{e^{\beta r\sum_{a=1}^n\s^a}}{[2\cosh(\beta r)]^n} ,
\label{eq:RS_parametriz_per_r}
\ee
where the density $R_n(r)$ is non-negative and normalized to $1$ in the limit $n\to 0$. In order to compute the $O(1/N)$ correction to the free energy, we need also to compute its normalization up to order $O(n)$. Inserting the parametrization \eqref{eq:RS_parametriz_per_r} in the equation defining $r(\s)$, i.e. Eq. \eqref{eq:Equzione_per_il_campo_r}, and considering also the $n$-dependence of the distribution $P_n(h)$ parametrizing $\rho(\s)$,  we obtain
\be
\int \dd r\ R_n(r)=1-n\,\mathbb{E}_{J,r,u}\log\left[
\frac{\cosh(\beta J)\cosh(\beta r + \beta u)}{\cosh(\beta r)\cosh(\beta u)}
\right]+O(n^2).
\label{corrR}
\ee
The random variable $u$ is called a cavity bias and is drawn from the distribution 
\be
Q(u) = \mathbb{E}_J\int \dd h\ P(h)\, \delta[u-\hat{u}(\beta,J,h)] ,
\ee
with $P(h)$ solution of Eq. \eqref{eq:cavityEq}. In Eq. \eqref{corrR} the random variable  $r$  is distributed as $R(r)=\lim_{n \to 0} R_n(r)$, solution to Eq. \eqref{eq:Equzione_per_il_campo_r}.

With this considerations in mind, the matrix $T(\s,\tau)$ can be written, for small $n$, as 
\be
T(\s,\tau) = \mathbb{E}_{J,r}\left[
\prod_{a=1}^n 
\exp\left(
\beta J\s_a\tau_a+\beta r\tau_a
\right)
\right] -\
n\,\mathbb{E}_{J,r,u}\log\left[
\frac{
2\cosh(\beta J)\cosh(\beta r +\beta u)
}
{
\cosh(\beta u)
}
\right] + O(n^2) .
\label{eq:Texpansion}
\ee
The first term in Eq.\eqref{eq:Texpansion} is the replicated transfer matrix of a $1$-dimensional disordered Ising chain with random couplings $J$ and random fields $r$. Let's call $T_n(\s,\tau)$ this first term.

The second term in Eq. \eqref{eq:Texpansion} is proportional to the thermodynamic free-energy density $\phi$ of an Ising chain with random couplings $J$ and random fields $r$ \cite{Monasson96, Lucibello2014}, explicitly:
\be
\mathbb{E}_{J,r,u}\log\left[
\frac{
2\cosh(\beta J)\cosh(\beta r +\beta u)
}
{
\cosh(\beta u)
}
\right] = -\beta\phi .
\ee
The full matrix $T(\s,\tau)$, in the limit $n\to 0$, then becomes:
\be
T(\s,\tau) = T_n(\s,\tau) + n\beta\phi + O(n^2) .
\ee
Taking the trace $\tr\left(T^\ell\right)$ we find
\be
\tr T^\ell = \tr T_n^\ell + n\ell\beta\phi + o(n^2) .
\ee
Now we observe that 
\be
\lim_{n\to0}\partial_n
\tr T_n^\ell = -\beta\phi_{\ell}^c ,
\ee
where $\phi_{\ell}^c$ is the free energy of a closed chain (loop) of length $\ell$, receiving a field $r$ on each of its vertex. Eventually taking the derivative and then the limit $n\to 0$ of the full trace $\tr T^\ell$, we get
\be
\lim_{n\to0}
\partial_n
\tr T^\ell = -\beta\left(
\phi_{\ell}^c - \ell\phi\right) \equiv \Delta\phi_\ell
 .
\label{eq:TracciaFinale}
\ee
Coming back to the equation \eqref{eq:DeltaFprimaDelLimite} for $f_1$, and substituting the previous result \eqref{eq:TracciaFinale}, we finally obtain the formula given in the main text:
\be
f_1 = \sum_{\ell=3}^{\infty}\frac{(c-1)^\ell}{2\ell} \Delta\phi_\ell .
\ee
\end{widetext}

\end{document}